\documentclass[twocolumn]{aastex631}
\usepackage{xcolor}
\usepackage{soul}

\begin{document}

\title{Fast-Cadence High-Contrast Imaging with Information Field Theory}

\author[0000-0002-8873-8215]{J. Roth}
\affiliation{Max Planck Institute for Astrophysics, Karl-Schwarzschildstr. 1, 85748 Garching, Germany}
\affiliation{Ludwig-Maximilians-Universit\"at, Geschwister-Scholl Platz 1, 80539 Munich, Germany}
\author[0000-0001-9539-2112]{G. Li Causi}
\affiliation{INAF - National Institute for Astrophysics Istituto di Astrofisica e Planetologia Spaziale, Via Fosso del Cavaliere 100, 00133 Roma, Italy}
\affiliation{INAF - National Institute for Astrophysics Osservatorio Astronomico di Roma, Via Frascati 33, 00078 Monte Porzio Catone (Rome),  Italy}
\author[0000-0003-1033-1340]{V. Testa}
\affiliation{INAF - National Institute for Astrophysics Osservatorio Astronomico di Roma, Via Frascati 33, 00078 Monte Porzio Catone (Rome),  Italy}
\author[0000-0001-5226-1171]{P. Arras}
\affiliation{Max Planck Institute for Astrophysics, Karl-Schwarzschildstr. 1, 85748 Garching, Germany}
\affiliation{Technische Universität München (TUM), Boltzmannstr. 3, 85748 Garching, Germany}
\author[0000-0001-5246-1624]{T. A. En{\ss}lin}
\affiliation{Max Planck Institute for Astrophysics, Karl-Schwarzschildstr. 1, 85748 Garching, Germany}
\affiliation{Ludwig-Maximilians-Universit\"at, Geschwister-Scholl Platz 1, 80539 Munich, Germany}

\begin{abstract}
Although many exoplanets have been indirectly detected over the last years, direct imaging of them with ground-based telescopes remains challenging. In the presence of atmospheric fluctuations, it is ambitious to resolve the high brightness contrasts at the small angular separation between the star and its potential partners. Post-processing of telescope images has become an essential tool to improve the resolvable contrast ratios. This paper contributes a post-processing algorithm for fast-cadence imaging, which deconvolves sequences of telescope images. The algorithm infers a Bayesian estimate of the astronomical object as well as the atmospheric optical path length, including its spatial and temporal structures. For this, we utilize physics-inspired models for the object, the atmosphere, and the telescope. The algorithm is computationally expensive but allows to resolve high contrast ratios despite short observation times and no field rotation. We test the performance of the algorithm with point-like companions synthetically injected into a real data set acquired with the SHARK-VIS pathfinder instrument at the LBT telescope. Sources with brightness ratios down to $6\cdot10^{-4}$ to the star are detected at $185$ mas separation with a short observation time of $0.6\,\text{s}$.
\end{abstract}

\section{Introduction}

\begin{figure*}[ht!]
\includegraphics[width=18.2cm]{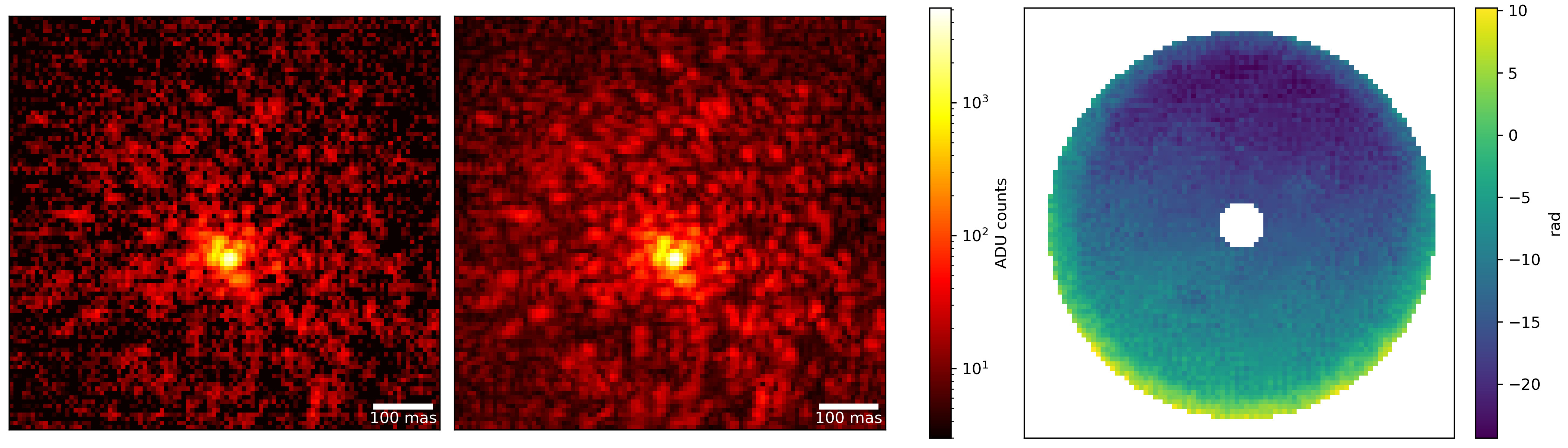}
\caption{Left panel: Instantaneous PSF in the focal plane (rebinned to $100\text{x}100$ pixels) of the $5.5$mag star Gliese 777, acquired with the SHARK-VIS Forerunner instrument at the LBT in a $1\,\text{ms}$ exposure: the interference pattern shaped by the atmospheric fluctuations (``speckles'') are well visible all around the star. Central panel: HCBI reconstruction of the PSF convolved star for the corresponding frame, shown with the same logarithmic colormap. Right panel: The reconstruction in the pupil plane of the optical path difference field for that frame of the Gliese 777 sequence, which determines the reconstructed PSF shown in the central panel. \label{fig:rct_data}}
\end{figure*}

The direct imaging of exoplanets is challenging due to the high brightness contrasts between a star and its potential exoplanets at angular separations of tiny fractions of an arcsecond, a context called ``high-contrast imaging'' (HCI; see \citet{Marois_2006}). The atmospheric turbulence that corrupts the incoming wavefront before entering the telescope, thus producing a distorted point spread function (PSF) at the focal plane, makes it difficult to resolve these high brightness contrasts. This corruption, especially pronounced for large telescopes, is nowadays mostly compensated by the extreme adaptive optics (ExAO) technology \citep{Esposito_2010}. However, the correction is not perfect, and the post-AO PSF still shows high spatio-temporal variations, with typical times of a few ms and scales of a few mas (see \citet{Stangalini_2017}). This results in a spread of the star light into a complex evolving pattern of spots, called ``speckles'', surrounding the star (see Fig.~\ref{fig:rct_data}, left panel) and hiding the much fainter companions.

To further improve the achievable contrast ratios, data post-processing techniques have become an essential part of HCI.
Mathematically, such data post-processing is a reconstruction problem in which a subtle signal shall be retrieved that is affected by three disturbances: i) a strong disturbing signal, i.e., the star's PSF with its photon noise, ii) its strong spatio-temporal variability, called ``speckle noise'', and iii) the signature and read-out noise of the detector.
Many post-processing algorithms have been developed for HCI, like the classical Angular Differential Imaging (ADI; \cite{Marois_2006}), the Speckle Free ADI (SFADI; \cite{LiCausi2017}), the Local Combination of Images (LOCI; \cite{Lafreniere_2007}), the Stochastic Speckle Discrimination (SSD; \cite{Walter_2019}), the Principal Component Analysis ADI (PCA-ADI; \cite{Soummer_2012}, \cite{Amara2012}),  as well as the recent NMF data imputation method (\cite{Ren_2020}). All of them estimate the time-dependent PSF leveraging on the large field rotation in a sequence of pupil-stabilized images across a long temporal interval. This model is then used to subtract the central star contribution from the data.

Here we present a different approach, which works on fast-cadence frames sequences to jointly reconstruct the static true brightness distribution of the source (i.e. star and companions) and the temporal evolution of the PSF. We use an explicit forward data modeling that simulates the natural information flow from the signal to the detector, as recently done by \cite{Hope2022} and, in a different way, by the MAYO (\cite{Pairet_2021}) and REXPACO (\cite{Flasseur_2021}) methods, and by the method proposed by \cite{Rodack21} and \cite{Frazin21}. Thereby we exploit the statistical properties of the fast cadence PSFs (whose exposure time is comparable to the speckles timescale).
In particular, we do this by using the information field theory (IFT; \citet{Ensslin_2019})

IFT uses Bayesian probability theory and methods from statistical field theories to infer fields, e.g., source distribution and PSF, from noisy data, allowing to incorporate prior knowledge from physical constraints.
In astronomy it has already been used in a number of contexts e.g. in radio astronomy (\citealt{Arras2021}, \citealt{Arras2019}, \citealt{Arras2022}), galactic dust tomography \citep{Leike2020}, or Faraday sky imaging \citep{Hutschenreuter2020}. This is the first time IFT is applied to HCI and hereafter we demonstrate its ability to reveal faint companions at one-tenth of an arcsecond separation from the host star in a very short acquisition time and without the need for field rotation. Since, at its core, our method rests on Bayesian inference, we name our algorithm High Contrast Bayesian Imaging (HCBI). Our code is open source and publicly available at: \url{https://gitlab.mpcdf.mpg.de/ift/public/hcbi}.

The paper is organized as follows. In Section~\ref{sec:IFT} we outline the general concepts and methods of IFT as needed for the HCI problem. In Section~\ref{sec:physics} we develop a physically inspired prior model for short exposure imaging. Section~\ref{sec:likelihood} outlines the symmetries in the likelihood. Section~\ref{sec:data} presents the results on real data, also in comparison with other methods. Finally Section~\ref{sec:conclusion} discusses the performance and planned improvements.

\section{Information Field Theory and its application to HCI}\label{sec:IFT}
IFT is a theoretical framework for the reconstruction of fields, i.e. continuous quantities having a value at any location, from measurement data. In the HCI context, examples for fields are the static source brightness distribution, which we call \textit{object}, and the time-variable PSF, which are both continuous in the sky or detector plane, respectively.

As fields are continuous quantities, they have an infinite number of degrees of freedom (d.o.f.) unless we restrict them to be of a parametric from. Thus reconstructing them non parametrically from finite measurement data is impossible without additional information. IFT provides the theoretical foundation to incorporate such necessary pieces of additional information into field reconstruction algorithms building on Bayesian inference.

Many physical fields vary mostly smoothly in space and time as the physical processes that govern them typically erase discontinuities quickly. Thus values of the field at nearby locations are correlated, and below a certain distance, no significant differences are expected. This effectively reduces the infinite number of d.o.f. to a finite number, enabling the inference of the field from finite data sets. Hence the key concept for inferring infinite-dimensional quantities is their correlation. Should the correlation structure be unknown a priori, it can be inferred simultaneously with the field as described in Section~\ref{sec:correl}. 
On a computer, these fields are sampled into pixels or voxels, requiring that the samplings are fine enough to resolve all relevant structures.
The open-source Python Bayesian field inference package NIFTy\footnote{\url{https://gitlab.mpcdf.mpg.de/ift/nifty}} (\cite{asclnifty5}, \citet{nifty3}, \cite{Selig2013}) is a numerical implementation of IFT, on which also the method presented in this paper builts on.

From the mathematical point of view, the HCI is described by the imaging equation, valid within an isoplanatic patch of the sky, for which the data vector $d=(d_1, d_2, ...)$, i.e. the pixel values of all acquired frames, are a convolution of the object $s$ with the PSF $p$ to which some noise $n$ is added,
\begin{equation}\label{eq:imaging}
    d_i(x) = (p * s)(x, t_i) + n_i
\end{equation}
where $x$ is the detector grid coordinate and $t_i$ the acquisition time of frame $i$. The PSF $p(x,t)$ includes the atmosphere and the optics response and depends on the coordinate $x$ and time $t$, while the object $s(x)$ is assumed to be static in time.

Our method computes from $d$ a Bayesian estimate of $s$ and $p$, i.e. it infers the posterior probability distribution $P(s,p|d)$ of $s$ and $p$ given the data $d$, following Bayes' Theorem:
\begin{equation}\label{eq:bayes}
    P(s,p|d) = \frac{P(d|s,p) P(s,p)}{P(d)}.
\end{equation}
Thereby the likelihood $P(d|s,p)$ is the probability of the data $d$ for a given realization of the object $s$ and the PSF~$p$; $P(s,p)$ is the prior, which encodes all previous knowledge about $s$ and $p$, like the correlation of $s$ and $p$, and physical constraints, like the strict positivity of those fields. Finally, the term $P(d)$ is the evidence, which only acts as a normalization constant that is not used in our algorithm.

In our case, we usually deal with frames sequences containing thousands of images, each having in the order of $10^4$ to $10^5$ spatio-temporal pixels, so that $s$, $p$, and $d$ are all very high dimensional quantities, and Eq.~\ref{eq:bayes} contains too many variables to be computed directly. Thus a fast and scalable algorithm is needed to infer an approximate posterior $P(s,p|d)$ and get a solution for $s$ and $p$.

\subsection{Metric Gaussian Variational Inference}\label{sec:MGVI}
The simplest approximation of the posterior is only to compute its maximum so that the solution for $s$ and $p$ corresponds to the highest probability density. This method is commonly referred to as maximum a posteriori (MAP) probability estimation.\\
Several more accurate algorithms for approximating the posterior distribution exist.
A suitable iterative algorithm for this work, capable of dealing with a very high dimensional parameters space, is Metric Gaussian Variational Inference (MGVI; \cite{Knollmueller2019}).
The concept of MGVI is to iteratively approximate the true posterior with a multi-dimensional Gaussian distribution. This Gaussian approximation is not directly performed on the parameters of interest, e.g. $p$ and $s$, but on some model parameters $\Theta$. The mappings $p(\Theta)$ and $s(\Theta)$ can be built according to physical considerations encoding prior knowledge on $s$ and $p$. In the next sections, these mappings are outlined in detail. Furthermore, performing the Gaussian approximation in a latent parameters space instead of directly on $p$ and $s$ enables the algorithm to capture the non-Gaussian statistics of $s$ and $p$. This more accurate approximation provides multiple advantages compared to the MAP estimator.

First, similar to MAP, MGVI has a near to linear scaling with the problem size, which makes it suitable for large problems with many free parameters.

Second, MGVI probes the phase space volume of the posterior, while MAP does not. This ensures that differences in the prior volume are more properly reflected in the inference with respect to MAP. This is of particular importance for our simultaneous object and PSF reconstruction since we are dealing with a degenerate inference problem in which several quantities are able to explain the same data feature. As only prior information (including its phase space volume)  can break many of these degeneracies, its more proper treatment by MGVI  is a clear advantage over MAP.

To summarize, IFT is a theoretical framework for Bayesian reconstructions of field-like quantities. MGVI is a specific numerical algorithm for obtaining an approximation of the actual posterior distribution for a given prior and likelihood model.

To apply MGVI to Eq.~\ref{eq:bayes} we need a model of the prior $P(s,p)$ and a model of the likelihood $P(d|s,p)$, that, in order to be meaningful, need to be derived using physically motivated arguments, as we are going to describe in Sec.~\ref{sec:physics} and Sec.~\ref{sec:likelihood} hereafter.

\section{Prior model}\label{sec:physics}
The MGVI algorithm requires the object prior and the measurement likelihood in the form of a generative model to draw realizations of $s(\Theta)$ and $p(\Theta)$ out of their probability distributions, starting from standard Gaussian distributed random variables $\Theta$. In the following, we build up such generative models for the object, the atmosphere, the telescope, and the measurement instrument by following physical considerations.

\subsection{Object model}\label{sec:object_model}
Bayes' Theorem quantifies how to update the prior knowledge with newly acquired data. Therefore our prior probability model should include all knowledge about the imaged object before evaluating the actual image data.

This means that the best generative model for the prior distribution is tailored to the source under study and will be different if our target is e.g. a star with a companion, a young stellar object with an accretion disk, a resolved object, or something else.

As in this work we focus our investigation on the detection of exoplanets or low mass stellar companions next to a central star, we need a point source model for both the star and its potential partners (which for the sake of simplicity we will call "planets" in the following, disregarding to their physical nature). We also know that the brightness of these sources and their locations in the sky do not vary across the short observation interval; therefore, our model object is time constant.

We focus our present work on the most common case where the star's apparent diameter is smaller than the pixel size so that the prior for the central star is a fixed point source at the center of the field of view. The only free stellar parameter of this model is the star intensity since telescope jittering and pointing errors that shift the star's apparent position in the image will be included in the PSF model. The intensity of the star must be positive and can vary on logarithmic scales, so we use a log-normal prior for it, which also ensures strict positivity in the reconstruction.

\begin{figure}
    \plotone{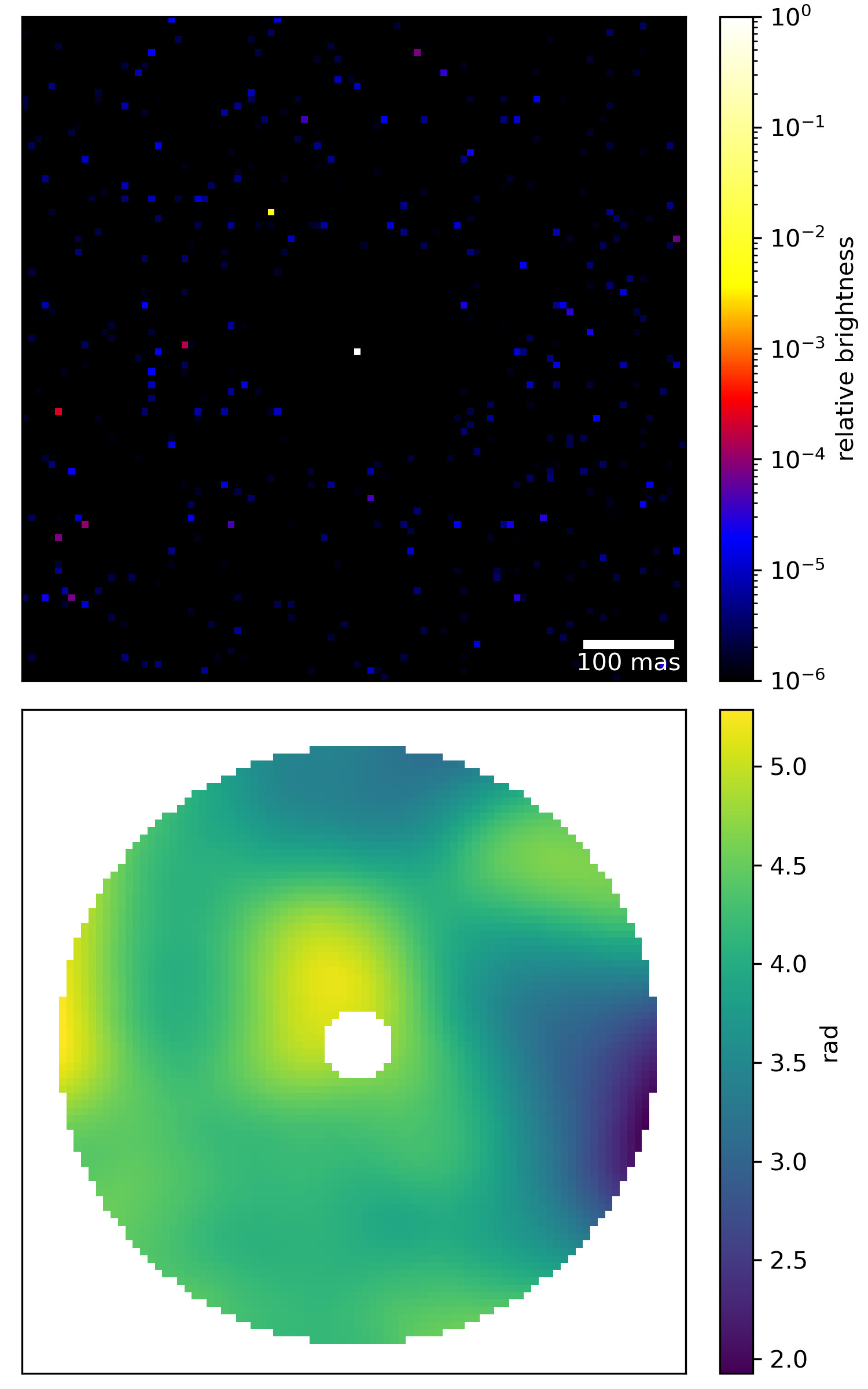}
    \caption{Samples drawn from the prior generative model. Top: One prior sample for the object, showing a random realization of the central star and the inverse gamma intensity distribution in all other pixels. In the close vicinity of the star it is enforced that the pixels have zero flux. This improves the quality of the reconstructed star PSF at the price of not being able to detect planets there. Bottom: One prior sample for the wavefront at telescope's pupil, generated from a Gaussian process with a spatial covariance structure given by the Matérn kernel of Eq.~\ref{eq:matern}. \label{fig:samples}}
\end{figure}

In contrast to the star, the locations of the planets are unknown, so we model them by inferring the intensity of every sky pixel, adopting for simplicity the same pixel grid of the data. We also know that there are no planets at most locations. Therefore the reconstructed intensity of most pixels should be very close to zero, given that the sky background has been subtracted from the data. To account for this statistics, we adopt independent inverse gamma priors for the brightness of any pixel. Consequently, the planet fluxes in the different pixels are a priori uncorrelated with each other, and it is much more probable for any pixel to belong to background than to host a planet.
A prior sample for the object model is displayed in Fig.~\ref{fig:samples}, top panel.

\subsection{PSF model}
Two separate effects contribute to a physical model for the evolving PSF. The first effect is the fluctuation in the atmospheric path length (Sec.~\ref{atmosphere}), while the second is the telescope optics (Sec.~\ref{telescope}). Because a typical high-contrast imaging observation has a field of view of the order of a few arcseconds, or even smaller, the assumption of isoplanatism of Eq.~\ref{eq:imaging} is valid, so we consider both contributions constant across the image.

\subsubsection{Atmospheric fluctuations model}\label{atmosphere}
The light from a pointlike astronomical source approaches the Earth as a plane wavefront but reaches the telescope's aperture as a distorted wavefront after corruption by the rapidly changing density fluctuations of the atmosphere. These distortions can be described by the optical path difference $\phi_{\text{atm}}(t,u)$, between the actual wavefront and the ideal plane wave.
It depends on time $t$ and position $u$ in the aperture plane. When expressed in units of wavelength, $u$ corresponds to spatial frequencies in the image plane. The instantaneous wavefront $\psi$ entering the aperture is thus given by:
\begin{equation}
    \psi_{\text{atm}}(t,u) = e^{i \phi_{\text{atm}}(t,u)}
\end{equation}

\subsubsection{Telescope model}\label{telescope}
The telescope's adaptive optics introduces opposite phase shifts to partly remove the path differences caused by the atmosphere. These phase shifts can be modeled in the same way as the atmosphere. Hence we have only one path length field $\phi_{\text{eff}}$ describing the combined effects of atmosphere and adaptive optics. 

Due to wind and mechanical vibrations, the telescope can jitter, introducing a shift in the detector's field of view. A shift in the detector space is equivalent to a pointwise multiplication of the Fourier transformed image by a phase gradient in the aperture space, with the slope of the gradient being proportional to the shift length. Therefore we add for both spatial directions a linear gradient $\phi_{\text{jit}}(t)$ with a time-dependent slope to the optical path difference.

The geometry of the telescope aperture blocks parts of this incoming wavefront, selecting the spatial frequencies that reach the focal plane. In our equation, we can model this with a multiplication factor $A(u)$, being zero for locations outside the aperture.

Finally, variations in the adaptive optics response and lensing effects in high layers of the atmosphere can change not only the phase but also the intensity of the wave. To also model such effects, we allow for a scalar time-dependent intensity factor, which modulates the amplitude in the aperture plane, which we write as an exponential parameter $e^{\alpha(t)}$ to guarantee the physical constraint of strict positivity.

To summarize, the resulting wavefront model in the optical pupil of the instrument can be written as:
\begin{equation}
    \psi(t,u) = A(u) e^{i \phi_{\text{eff}}(t,u) + i \phi_{\text{jit}}(t) + \alpha(t)}
\end{equation}

The final instantaneous image of the PSF model point source produced at the focal plane is the squared amplitude of the Fourier transform of the wave in the optical pupil:
\begin{equation}\label{eq:psf}
    p(t, x) = \left|\mathcal{F}_u^x \psi(t,u) \right|^2 = \left|\mathcal{F}_u^x A(u) e^{i \phi_{\text{eff}}(t,u) + i \phi_{\text{jit}}(t) + \alpha(t)} \right|^2.
\end{equation}

To this equation, we could add the dependence on $\lambda$, since the effects of atmosphere, adaptive optics, and telescope are wavelength-dependent. In this work, though, we do not make this explicit because we apply our method to narrow-band observations.

\subsubsection{Correlation prior}\label{sec:correl}
In every space-time voxel, $\phi_{\text{eff}}(t,u)$, $\phi_{\text{jit}}(t)$, and $\alpha(t)$ can have different values. Nevertheless, because of the physical mechanisms involved, not all values are equally likely. For example, the path length field $\phi_{\text{eff}}(t,u)$ will evolve smoothly in time and space since also the real atmosphere evolves smoothly. Similar, $\phi_{\text{jit}}(t)$ and $\alpha(t)$ will be correlated in time.

Modeling the correlations and using them as prior information substantially improves the reconstructions. Especially for the path length $\phi_{\text{eff}}(t,u)$, the correlations are important to deal with the degeneracy between object and PSF in the imaging equation (\ref{eq:imaging}). In Section~\ref{sec:data}, we demonstrate the benefits of accurately reconstructing these correlations.

We model all correlated fields as zero-centered Gaussian processes, thanks to the already cited standard variables $\Theta$. The correlation structures of $\phi_{\text{eff}}(t,u)$, i.e. its power spectrum, might be different in spatial and temporal directions. Because we assume the spatial correlations not to change significantly during an observation, we factorize the Gaussian process into a temporal and a spatial axis and write the total correlation kernel as the outer product of spatial and temporal kernels. For both axes we employ a Mat\'{e}rn kernel (\cite{Genton2002}, \citet{Matern1986}). The Mat\'{e}rn covariance is translation invariant and therefore diagonal in Fourier space. The Mat\'{e}rn kernel can be parametrized in Fourier space
\begin{equation}\label{eq:matern}
    \mathcal{P}(k) = \frac{a^2}{(1 + (k / b)^2)^c},
\end{equation}
where $\mathcal{P}(k)$ indicates the power spectrum with $k$ as frequency.

The parameters $a$, $b$, and $c$ are a priori unknown and are simultaneously reconstructed with the path length field. The parameter $a$ determines the amplitude of the fluctuations, $b$ is a characteristic length scale, and $c$ is the spectral slope of the spatial or temporal fluctuations. For the temporal and spatial axis, the parameters of the kernel can be different. 

For $\phi_{\text{jit}}(t)$ and $\alpha(t)$ a similar, but non-parametric, correlation model is used, whose details can be found in \cite{Arras2022}.

Fig.~\ref{fig:samples}, bottom panel depicts one realization of the wavefront field $\phi(t,u)$ drawn from the described generative model, showing its spatial correlations.

\section{Likelihood model}\label{sec:likelihood}
For a given object $s$ and evolving PSF $p$, the model described above determines the time dependent intensity field in the focal plane. The detector's readout and the photon count statistics are accounted for in Eq.~\ref{eq:bayes} by the likelihood term $P(d|s,p)$ by using Poisson and Gaussian statistics, respectively.

For simplicity, the details of the detector signature like pixel-to-pixel bias non-uniformities, frame-to-frame column bias variations, and pixel-to-pixel non-uniform responses are currently not included in the forward model but will be incorporated in the future. For the current work, the images are pre-corrected for these effects before the actual reconstruction (see Sec.~\ref{sec:data}).

\subsection{Symmetries of the likelihood}\label{subsec:sym}
A given PSF $p(t,x)$ does not uniquely determine the wavefront $\psi(t,u)$, because several distinct path length functions will result in the same PSF. In other words, the PSF is symmetric under specific transformations of the wavefront. A detailed review of these symmetries can be found in \cite{Paxman2019}, here we shortly outline them:
\begin{itemize}
    \item Global phase offset symmetry: A global additive offset to the path length does not change the PSF, since the absolute value in Eq.~\ref{eq:psf} remains constant.
    \item $2\pi$ ambiguity: One can locally add multiples of $2\pi$ to the path length field without changing the PSF, since $e^{i 2\pi n}=1$ for all $n \in \mathbb{Z}$.
    \item  Sign ambiguity of the symmetric part of the wavefront (for symmetric apertures): If one writes the path length field $\phi(t,u)$ as the sum of a symmetric part $\phi_s(t,u) = \phi(t,u) + \phi(t,-u)$ and an anti-symmetric one $\phi_a(t,u) = \phi(t,u) - \phi(t,-u)$, then $\phi_+(t,u) = \phi_a(t,u) + \phi_s(t,u)$ and $\phi_-(t,u) = \phi_a(t,u) - \phi_s(t,u)$ both correspond to the same PSF. 
\end{itemize}
Since the PSF model has these three symmetries also the likelihood has these symmetries because the probability of a given frame in the data is the same for any realization of these symmetries in the wavefront.

The symmetries are divided into two categories, continuous symmetries and discrete symmetries. The first symmetry is continuous, while the second and third are discrete. For the continuous symmetry, one solution of the phase screen can be transformed into another solution by a series of local changes without ever changing the resulting PSF or the likelihood. This is not possible for the two discrete symmetries. For example, to transform the path length function $\phi_+(t,u)$ into $\phi_-(t,u)$ without changing the PSF, one has to entirely change the sign of the symmetric part.

MGVI, our inference scheme, iteratively improves the current estimate by making local changes. After performing a sequence of local optimization steps, it can happen that parts of the path length function converged to $\phi_+(t,u)$ and other parts to $\phi_-(t,u)$. Similarly, parts of the path length function can have phase offsets of multiples of $2 \pi$. This can be suboptimal for exploiting the correlations of the path length function since $2\pi$ jumps and jumps between $\phi_+(t,u)$ and $\phi_-(t,u)$ make the path length function appear less correlated. Even in the presence of some artefacts, there are correlations in the path length function, which can help to improve the reconstruction of the sky brightness.

\subsection{Double path length function model}

As outlined in the previous section, the phase wrap artefacts caused by the discrete symmetries of the likelihood are only hardly removed by the MGVI optimization. To reduce these artefacts, we alter the model for the initial iterations of the optimization in order to separate the smooth structures from the artefacts. We add an auxiliary path length function for the initial iterations and superimpose the resulting image in the focal plane with the image determined by the primary path length function. Thus each feature in the data can be explained by either of the two path length functions.

\begin{figure*}[!h]
    \includegraphics[width=18cm]{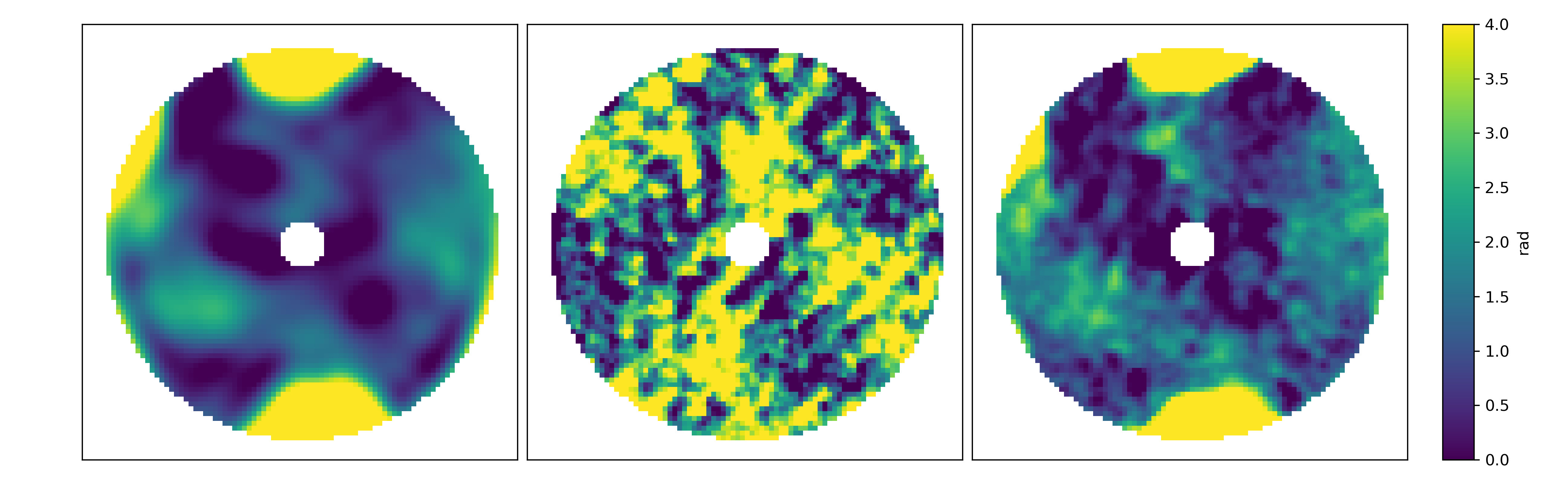}
    \caption{Path length functions in the pupil plane at different stages of the optimization. The left panel shows the primary path length function before the auxiliary path length function was removed. As described in the main text it only contains large scale structure explaining the central peak of the PSF. The central panel depicts the auxiliary path length function at the same stage of the optimization. In the right panel the final result of the path length reconstruction is plotted. Many small scale structures form the auxiliary path length function have been transferred to the primary path length function without introducing phase wrapping effects. \label{fig:two_path_length}}
\end{figure*}

For the auxiliary path length function, we tune the hyper-parameters of the Mat\'{e}rn kernel towards favoring flatter power spectrum slopes, thus allowing less correlation for this function. When starting the inference, the additional path length function will quickly accumulate many small-scale structures at the cost of having multiple symmetry artefacts. In contrast, the primary path length function will be smooth, mainly explaining the central peak of the PSF without modeling the small-scale features. Thus the primary path length function reconstruction will be physically plausible, having only little artefacts, but lacking small-scale structures. In Fig.~\ref{fig:two_path_length} (left and central panel) the primary and auxiliary path length functions are depicted for an intermediate stage of the optimization.

In subsequent iterations, we reduce the weight of the image resulting from the auxiliary path length function in the superposition with the primary image. This forces the reconstruction to explain an increasing amount of small-scale features by the primary path length function, thus gradually transferring small-scale features back to the main path length function without creating phase wrap artifacts. In the right panel of Fig.~\ref{fig:two_path_length} the path length function is plotted after the transfer of the small scale features from the auxiliary path length function.

\begin{figure*}[!h]
    \includegraphics[width=18cm]{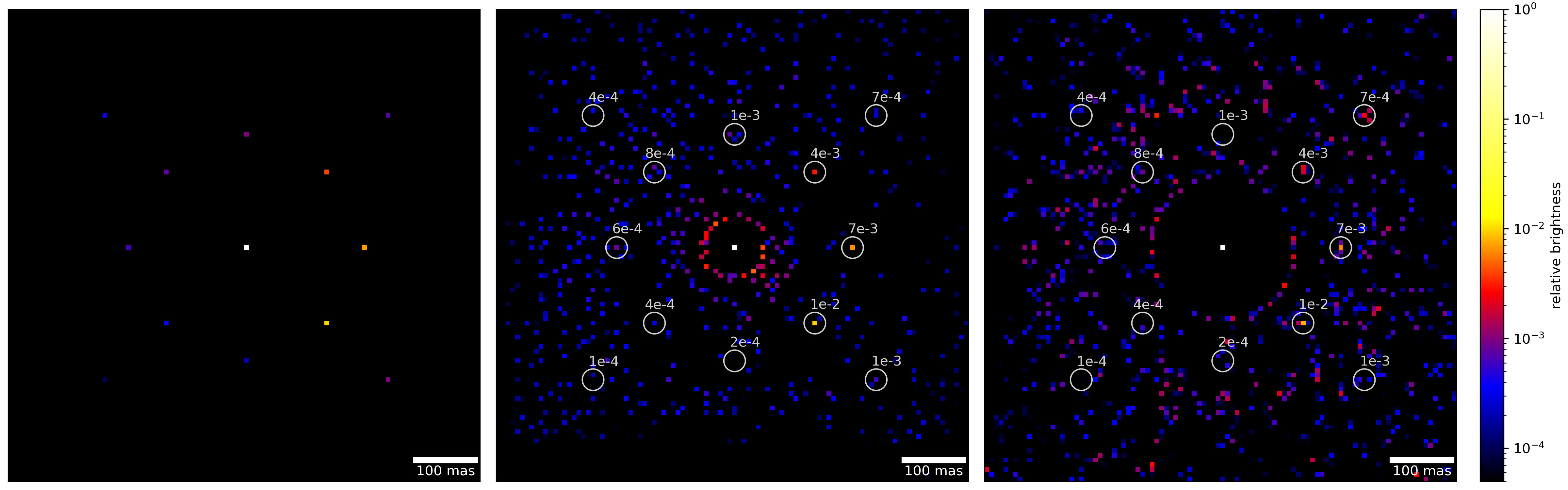}
    \caption{Left: Ground truth object: A central star surrounded by 12 injected planets with contrasts ranging from $10^{-2}$ to $10^{-4}$, shown in the field of view of the frames. Center: HCBI object reconstruction from $0.6\,\text{s}$ of Gliese 777 data from $600$ sequential frames. Right: HCBI object reconstruction from using only $0.1\,\text{s}$ of the data. The $0.1\,\text{s}$ reconstruction can only detect the brightest 3 companions. Furthermore the $0.6\,\text{s}$ reconstruction has a significantly lower background residual. \label{fig:rct_obj}}
\end{figure*}

\section{Application to a real data}\label{sec:data}
We applied the described algorithm to real on-sky frames sequence acquired with the SHARK-VIS pathfinder instrument, called Forerunner \citep{Pedichini_2017}. SHARK-VIS \citep{Mattioli_2018} is the upcoming visible band fast imaging camera for high-contrast imaging at the 8-meter class Large Binocular Telescope (LBT).

The data were acquired on the $5.5\,\text{mag}$ star Gliese 777, imaged at a frame rate of $1\,\text{KHz}$ through a $40\,\text{nm}$ FWHM filter centered at $630\,\text{nm}$, and recorded by an Andor Zyla sCMOS camera. The low read noise of this camera, of less than $2$ electrons per pixel, and its small pixels of $6.5\mu m$ allows to use an image scale of $3.73\,\text{mas/pix}$, and a frame exposure of $1\,\text{ms}$, that was chosen to oversample the typical spatial and temporal scales of the speckles in the visible band, as measured in \citet{Stangalini_2017}. 

During the observation, the seeing was around $1\,\text{arcsec}$ FWHM, and the LBT adaptive optics system \citep{Esposito_2010} was correcting $500$ modes in closed loop. An exemplary frame is shown in Fig.~\ref{fig:rct_data}, left panel.

Fast cadence imaging allows for numerical re-centering of the frames and a frame selection discarding the few frames whose peak is fragmented by strong turbulence. This potentially yields two important advantages. First, it improves the planet’s detection due to the increased flux concentration, and second, it increases spatial resolution.\\
As described in section \ref{sec:physics} the presented method reconstructs the evolving instantaneous PSF, which makes it suitable to utilize the full potential of such fast-cadence observations.

The whole Gliese 777 data set consists of $1.2\cdot10^{6}$ sequential frames for an interval of $20$ minutes, acquired without using the telescope de-rotator (pupil-stabilized imaging) to get the necessary field rotation needed for the application of the standard angular differential techniques mentioned in the introduction.
These raw frames were pre-processed before the HCBI reconstruction to reduce detector artifacts, like the frame-to-frame column bias variations and the pixel-to-pixel non-uniform response.

For testing the HCBI reconstruction we only use $600$ sequential frames, corresponding to the first $0.6\,\text{s}$ of acquisition, in which we injected $12$ synthetic planets between $185\,\text{mas}$ and $315\,\text{mas}$ from the star and with contrasts ranging from $10^{-2}$ down to $10^{-4}$ (Fig.~\ref{fig:rct_obj}, left panel).

The mock planets are produced by shifting and scaling each frame to the position and intensity of the planets and adding the corresponding Poisson random counts to the data. These frames have then been binned to a plate scale of $7.46\,\text{mas/pix}$ in order to work with $100\times100$ pixel frames and reduce the high computational workload of the processing.

The MGVI inference process for the current algorithm starts by generating a set of random prior samples from all input variables of the generative model (Sec.~\ref{sec:physics}), e.g. the object, the optical path length, the telescope jittering, and the scale of the wavefront amplitude.
Afterwards the algorithm computes a model of the frames by a convolution of the thereby generated object samples with the PSF samples and evaluates the likelihood (Sec.~\ref{sec:likelihood}).

Then the MGVI algorithm optimizes the Gaussian posterior approximation with these samples. In the following, this procedure is iterated until convergence, using samples from the current estimate of the posterior instead of prior samples. During these iterations, the weight of the auxiliary PSF is reduced until only the primary PSF is left.

\begin{figure}[t]
    \plotone{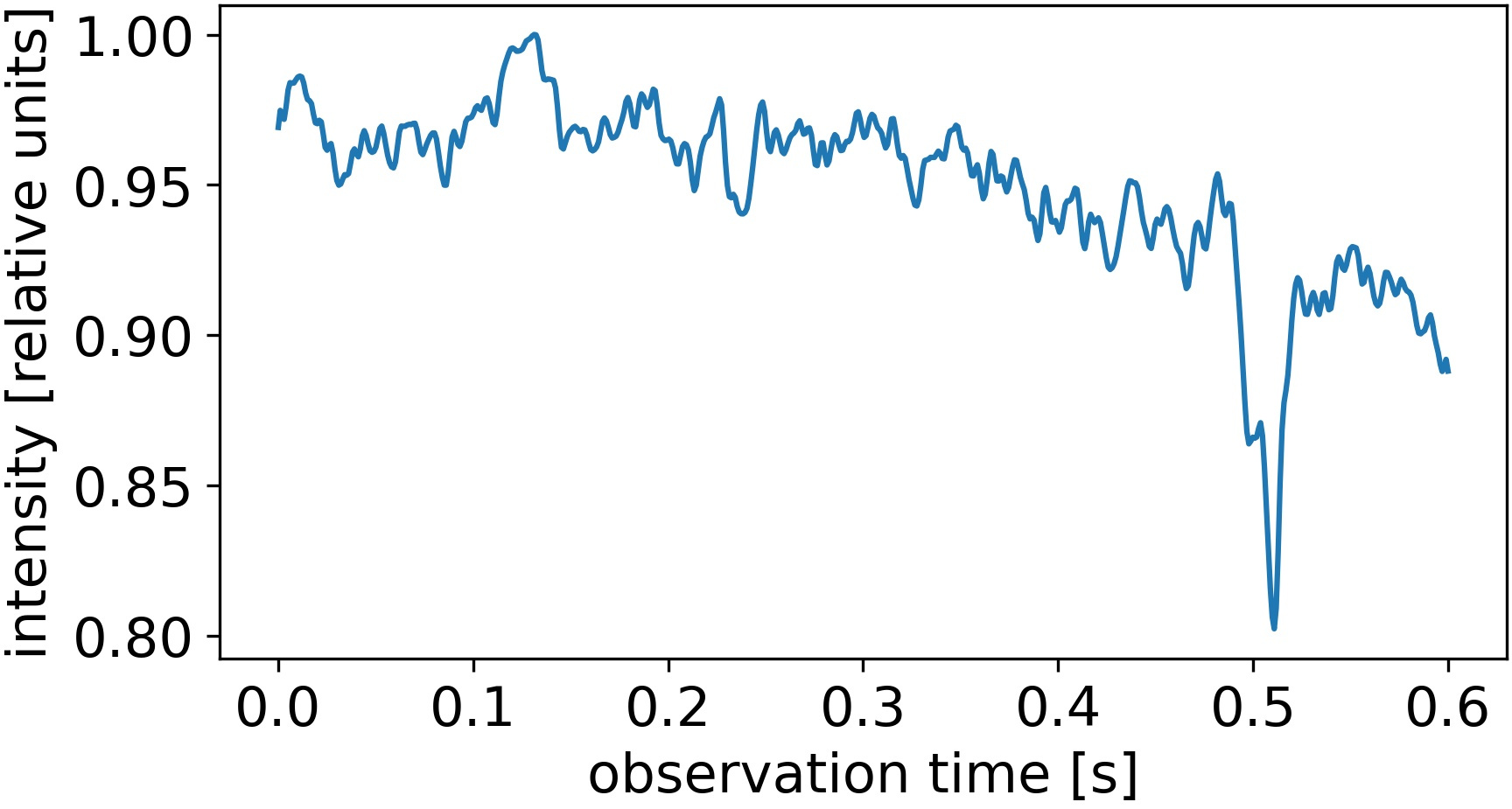}
    \caption{Reconstructed amplitude variation of the PSF $\exp(\alpha(t))$ as a function of observation time.  \label{fig:scale}}
\end{figure}

Finally, the sample average can be computed for all quantities in the generative model (Sec.~\ref{sec:physics}). Thus HCBI does not only reconstruct the astronomical object and the optical path length but simultaneously calibrates for all the modeled effects, like e.g. the total flux variation $\exp(\alpha(t))$ due to atmospheric scintillation, as shown in Fig.~\ref{fig:scale}.

Fig.~\ref{fig:rct_data}, in central and right panels, shows the model frame and the model wavefront for one data frame. All prominent data features are accurately reconstructed, indicating that the model is sufficiently flexible to describe the data. Smaller structures are attributed to noise and are not picked up in the reconstruction thanks to the prior favoring smooth path length functions. A strong correlation is noticeable in the wavefront, as expected from our physical considerations in Sec.~\ref{sec:physics}, and also following the prior model.

Fig.~\ref{fig:rct_obj}, central panel, depicts the reconstructed object after convergence of the algorithm.
It is composed of the pointlike star at the center, the correctly reconstructed planets, and a background residual mainly consisting of isolated points between $5\cdot10^{-3}$ and $10^{-4}$ intensity with respect to the star. The planets are all correctly reconstructed as pointlike objects in the right locations down to a contrast of $6\cdot10^{-4}$.

The residual background is not Gaussian, as we expect from the inverse gamma prior that we adopted for the source field.

This result demonstrates that a $0.6s$ acquisition is enough for the HCBI to achieve a detection limit of $6\cdot10^{-4}$ on a $5.5\,\text{mag}$ star without the need for field rotation. The measured intensities of the reconstructed planets are correct within an error of $5\%$ for the brightest and $17\%$ for the faintest with respect to the ground truth, while the identified planets less or equal to $10^{-3}$ are comparable with the false positives of the same level in the residual background.

It is worth noting that with fast-cadence imaging, the problem of quasi-static speckles that can mimic a planet signal in long exposure frames is much less severe. Instead, our false positives mostly come from noise.\\

\begin{figure*}[!h]
    \includegraphics[width=18.2cm]{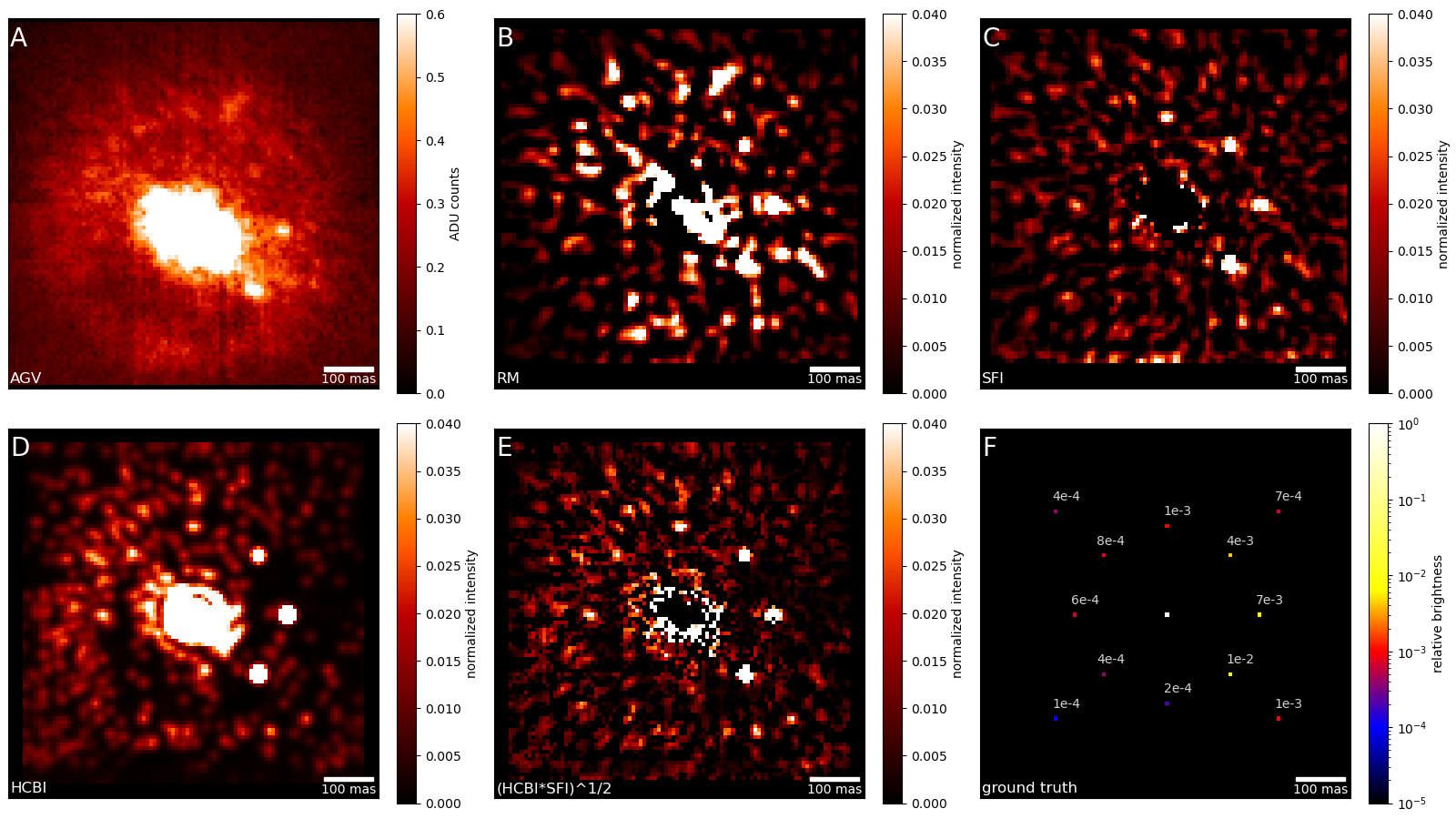}
    \caption{Performances of different methods for the same $600$ frames: a) simple average after aligning; b) average of the original frames after subtraction of a radial profile and a local median; c) multi-threshold SFI; d) HCBI reconstruction (namely, central panel of Fig.~\ref{fig:rct_obj}) convolved with a Gaussian of the same FWHM of the instrumental PSF; e) geometric average of results from SFI and HCBI; f) ground truth inserted into the data set. All the panels from b to e are shown with the same linear color scale in order to facilitate the comparison of the residuals.}\label{fig:comparison}
\end{figure*}

\subsection{Comparison with other methods}\label{sec:pca}
In order to directly compare the HCBI result with other methods, we convolved the object reconstruction with a Gaussian of the same FWHM of the instrumental PSF. This produces the image in Fig.~\ref{fig:comparison}, panel d.

Since field rotation is negligible on a $0.6\,\text{s}$ short sequence, the common angular differential methods cannot be used to process these frames. So in Fig.~\ref{fig:comparison} we made a comparison with i) the simple average of the frames after co-registration (panel a), ii) the simple star removal obtained by subtraction of average radial profile and local median (panel b), and iii) the result of a multi-threshold Speckle Free Imaging (SFI; \citet{ LiCausi2017}, \citet{Mattioli_2019}), shown in panel c.

We see that the reconstruction from HCBI (panel d) unambiguously reveals all planets down to $6\cdot10^{-4}$ (indicated for reference in panel f) against a fainter background residual than the SFI method, in which they are not distinguishable from the residual spots. Anyway, these planets are present in the SFI result, which suggests that a combination of the two results could even increase the detectability, since planets are correlated while residuals are not, as shown by their geometric average (panel e).

Such comparison, however, is only possible for sufficiently bright stars, because the SFI method is only efficient when speckles are well detectable in a single frame (\citet{ LiCausi2017}), while HCBI is less affected by the star being dim because it exploits the entire information across the whole frames sequence.

It is worth noticing that for applying any ADI-based technique, like e.g., the PCA-ADI, to these fast cadence data, at least $10^5$ frames would be needed for having a sufficient field rotation to avoid planet self-subtraction.

\section{Discussion}\label{sec:conclusion}
This paper describes a novel data modeling algorithm for obtaining direct high-contrast imaging of extrasolar planets, or faint stellar companions, at sub-arcsecond distance from their host star.
The algorithm is based on the mathematical methods of information field theory (IFT) and uses Bayesian inference to simultaneously reconstruct the true object and the PSF evolution from a kHz-rate frames sequence acquisition. For the reconstruction, it uses a set of prior information reflecting the physics of the object, the atmospheric perturbations, the telescope and detector response.
The prior model resembles the observational effects present at fast-cadence imaging instruments such as SHARK-VIS, allowing the method to unravel the full potential of fast-cadence frame sequences.
Special care is taken to deal with symmetries in the wavefront that are observationally indistinguishable but may affect the inference. The wavefront is modeled with a Gaussian process whose correlation power spectrum is learned from data as part of the inference process. Metric Gaussian Variational Inference is adopted at the core of the present algorithm to get robust convergence to an optimal Gaussian approximation of the posterior distribution, from which the final reconstruction is computed.

The performance of HCBI is tested on real on-sky data with artificially injected faint planets, and its detection limits are discussed in comparison with other methods. We show that the HCBI algorithm outperforms the Speckle-Free Imaging (SFI), producing a smaller residual and correspondingly increasing the planets detection limit. Moreover, in contrast to SFI, which needs to detect speckles in each frame, the HCBI method is not limited to bright stars because it uses the whole data cube at once. Finally it does not need field rotation, which is needed by the ADI-based methods to avoid self-subtraction of the planet.

To study the sensitivity increase of the HBCI method as more frames are processed, we also reconstructed a subset of the data consisting of only $100$ frames corresponding to $0.1\,\text{s}$ of observation. The result of this $100$ frames reconstruction is depicted in Fig. \ref{fig:rct_obj} right panel. As expected, the sensitivity increases as more frames are processed.
Hence processing thousands of frames should improve detection limits further, with reduced background residuals and less false positives. In practice however, we are currently limited to work with a few hundreds of frames, due to the high computational cost of the MGVI reconstruction, which took about $2$ weeks to converge to the $600$ frames result of Fig.~\ref{fig:rct_obj}, using two $2.4$GHz cores of an Intel Xeon Gold 6148 processor with a $20\,\text{GB}$ RAM. To face this problem, our plan for the continuation of this work is to study the possibility of splitting large datasets into a number of smaller chunks, to be treated independently and processed in parallel by adopting cloud computing, whose reconstructed information could be joined incrementally. This is physically justified because the whole speckles pattern becomes fully uncorrelated after $\sim 70-90\,\text{ms}$, as measured by \citet{Stangalini_2017}.

We expect to improve the performance of the method by including the detector signature into the generative data model, which in this work has been corrected by a preprocessing step. Incorporating the signature into the generative model would allow for a joint reconstruction of the object, the PSF evolution, and the detector fluctuations. Also, we will introduce more priors, specific for the cases of resolved stars or extended sources like circumstellar planet-forming disks, in order to enlarge the applicability of the method to all the most common high-contrast imaging situations.

Finally, we plan to ascertain the possibility of modeling long exposure frames, which requires a completely different PSF prior model. Possible models are under investigation but go beyond the scope of the present work.

\bibliographystyle{aasjournal}
\bibliography{references}{}
\end{document}